\documentclass[aps,twocolumn]{revtex4}
\usepackage{natbib}

\input epsf
\def\plotone#1{\centering \leavevmode
\epsfxsize= 1.0\columnwidth \epsfbox{#1}}
\def\be{\begin{equation}}
\def\ee{\end{equation}}
\def\bea{\begin{eqnarray}}
\def\eea{\end{eqnarray}}

\def\cmm2{{\,\rm cm^{-2}}}
\def\cm2{{\,{\rm cm}^2}}
\def\cmm3{{\,{\rm cm}^{-3}}}
\def\gcmm3{{\,{\rm g\,cm^{-3}}}}

\def\fun#1#2{\lower3.6pt\vbox{\baselineskip0pt\lineskip.9pt
  \ialign{$\mathsurround=0pt#1\hfil##\hfil$\crcr#2\crcr\sim\crcr}}}

\def\p3m{P$^3$M}

\def\fun#1#2{\lower3.6pt\vbox{\baselineskip0pt\lineskip.9pt
  \ialign{$\mathsurround=0pt#1\hfil##\hfil$\crcr#2\crcr\sim\crcr}}}

\begin{document}
\bibliographystyle{prsty}
\title{Comparing and combining Wilkinson Microwave Anisotropy (WMAP) probe results and Large Scale Structure}
\author{Licia \ Verde\footnote{The WMAP science team composed by: C. Barnes, C. Bennett (PI), M. Halpern, R. Hill, G. Hinshaw, N. Jarosik, A. Kogut, E. Komatsu, M. Limon, S. Meyer, N. Odegard, L. Page, H. Peiris, D. Spergel, G. Tucker, L. Verde, J. Weiland, E. Wollack, E. Wright. }}
\affiliation{Dept. of Astrophysical Sciences\\
Princeton University, Princeton, USA}
\date{\today}

\begin{abstract}
Since this is a meeting on inflation the title of this talk should probably have been ``The cosmos as a lab for inflation''. I will illustrate how, by combining WMAP  observations of the cosmic microwave background (CMB) with smaller scales CMB experiments and large-scale structure surveys, it is possible to constrain inflationary models. WMAP data in combination with these external data sets, offers a window to the very early universe and the physics of inflation:  
enables us  to confirm the basic tenets of the inflationary  paradigm and begin to quantitatively test inflationary models.  Understanding the complex physics and complicated systematics of the low redshift observations is crucial to make progress in this direction.

\end{abstract}
  \maketitle

\section{Introduction}
The first year results from the Wilkinson Microwave Anisotropy probe (WMAP) were  announced on February 11 2003. On the same day the telescope was renamed in honor of Prof. David Wilkinson, member of the science team and pioneer in the study of cosmic microwave background (CMB) radiation.
The primary goal of the WMAP mission  is to produce a high-fidelity all-sky polarization-sensitive  map of the cosmic microwave background (CMB) radiation to determine the cosmology of our Universe.
After a year of observation it has produced a full sky map of the microwave sky in 5 frequencies, with a resolution a factor 30 higher than the previous full sky map as produced by the COBE satellite \cite{Bennettetal96}.
This is the cleanest picture of the early Universe; the structures on the CMB --the pattern of hot and cold spots-- carry information about the composition, geometry,  age  etc. of our Universe (e.g., \cite{Jungmanetal96a,Jungmanetal96b}).
Here I will illustrate how it is possible to  constrain inflationary models by combining WMAP's CMB observations with external data sets: in particular, the smaller scales CMB experiments CBI and ACBAR, and observations of  large-scale structure probes such as the power spectrum of Two-degree-field galaxy redshift survey (2dFGRS; \cite{Collessetal01}) and the  linear matter power spectrum as recovered by \cite{Croftetal02} and \cite{Gnedinhamilton02} from Ly${\alpha}$ forest observations. We also consider the Type 1A supernoave results \cite{Riessetal01} and HST Key project Hubble constant determinations \cite{Freedmanetal01}.  The results of this analysis were reported in \cite{ Verdeetal03, Spergeletal03, Peirisetal03}.


\section{WMAP}

If the CMB is Gaussian (see \cite{Komatsuetal03}) then the  information in the mega-pixel maps can be compressed in the power spectrum \cite{Hinshawetal03}.
We show the WMAP temperature  power spectrum in Fig. 1 (left), where the points with  error-bars  are the (band power) data with the error given by the instrumental noise. The solid line is our best fit model and the gray area shows the cosmic variance uncertainty.
At the moment we can reliably measure only the temperature-polarization (TE) cross-correlation  power spectrum which is shown in Fig. 1 (right) where the points with error-bars are the band-power data.
The temperature power spectrum makes precise predictions for the TE power spectrum on large scales. The solid line in Fig. 1 (right) is this prediction from the best fit to the TT data only. For $\ell>10$ the data are in remarkable agreement with the TT predictions, this is a triumph for the standard cosmological model (see \cite{Peirisetal03}); the WMAP detection of TE anti-correlation at $\ell \sim 50-150$ rules out a broad class of active models and implies the existence of super-horizon adiabatic fluctuations at decoupling: both inflation and the Ekpyrotic scenario predict the existence of super horizon fluctuations.

For $\ell<10$ we detect a signal that cannot be seen in the TT data alone: this is the signal of early re-ionization \cite{Kogutetal03}. Such a strong  signal at low multipoles gives a  constraint on the integrated optical depth to the last scattering surface $\tau=0.17\pm 0.04$, for instantaneous reionization this corresponds to a reionization redshift of $z_{re}=17\pm 3$.

Why WMAP has such a good signal to noise?
A part from the fact that it is an all sky experiment which observed the sky for a full year, this is the result of an obsession of all those involved in designing and building the satellite for minimizing systematics and maximizing the signal to noise. 
During the designing and building  of the satellite the emphasis was on minimizing systematic errors so that the data could be straightforwardly interpreted. 
For example there are 10 different channels from 20 to 95 GHz, so that foreground subtraction is not a problem \cite{Bennettetal03,Hinshawetal03}.
Undesirable $1/f$ noise is minimized by the design of  WMAP radiometers \cite{Jarosiketal03}, uncorrelated noise between pixels, accurate determination of the beam patterns \cite{Pageetal03,Barnesetal03} 
and well understood properties of the radiometers are invaluable for this analysis.

The rest of the analysis must therefore have the same level of ``obsession''.
For example, to extract as much information as possible from the maps without introducing systematics, the TT power spectrum is obtained by optimally combining ``only'' 28 cross-channels correlations. This method has then be checked against another 2 independent methods \cite{Hinshawetal03}. 
Also, the point source subtraction has been done in 3 different ways \cite{Hinshawetal03, Bennettetal03, Komatsuetal03}(they all agree!). 
Finally in the covariance matrix we propagate all the effects we know about (beam errors, noise, sky cut etc.)
To cut a long story short, we spent a long time worrying about 2\% error on the error! 
There are effects that we are aware of, which we neglect; these effects 
give about 0.5 to 1\% error on the error.

\begin{figure}[h]
\begin{center}
\setlength{\unitlength}{1mm}
\begin{picture}(90,98)
\includegraphics{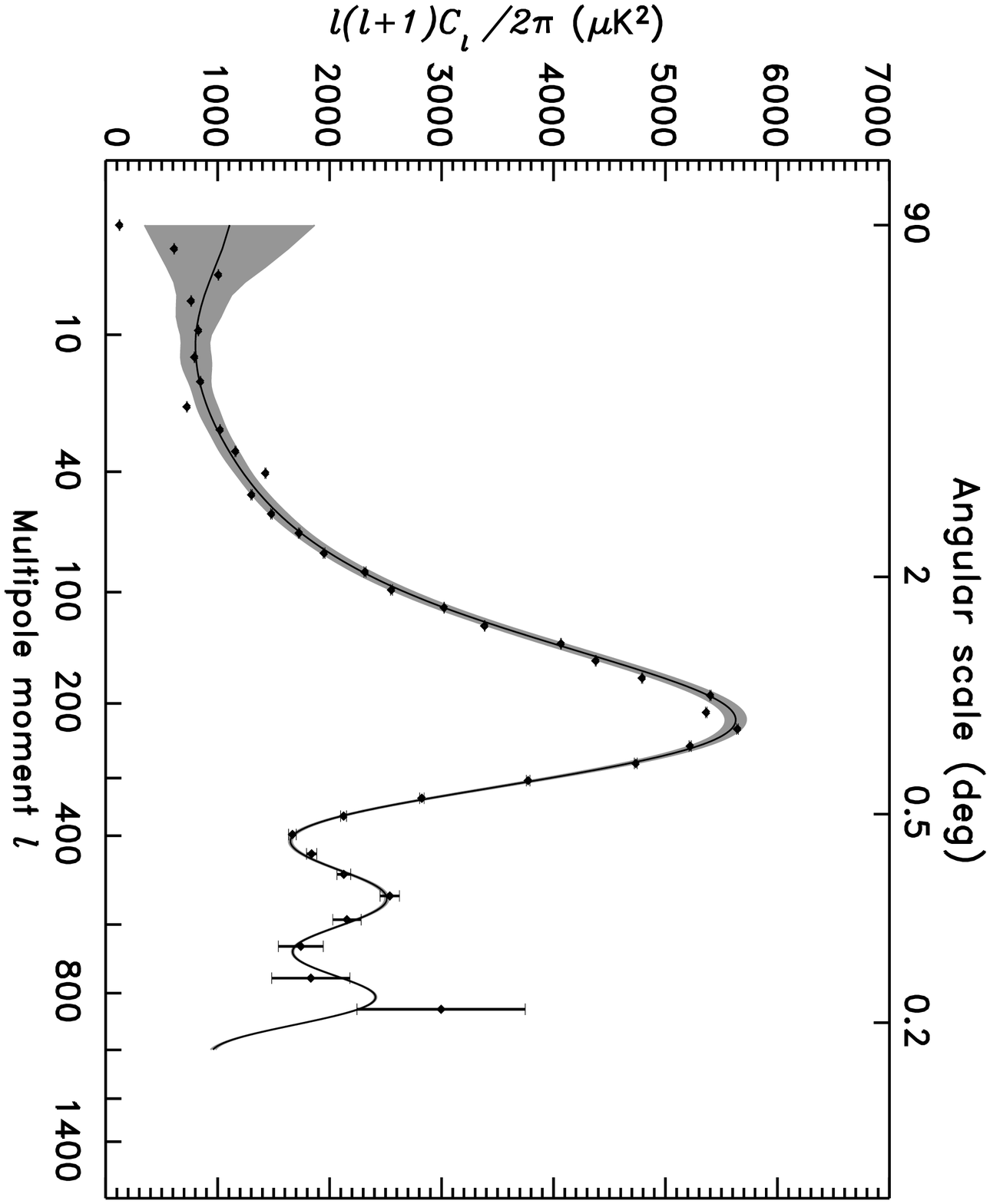}
\begin{picture}(90,30)
\includegraphics{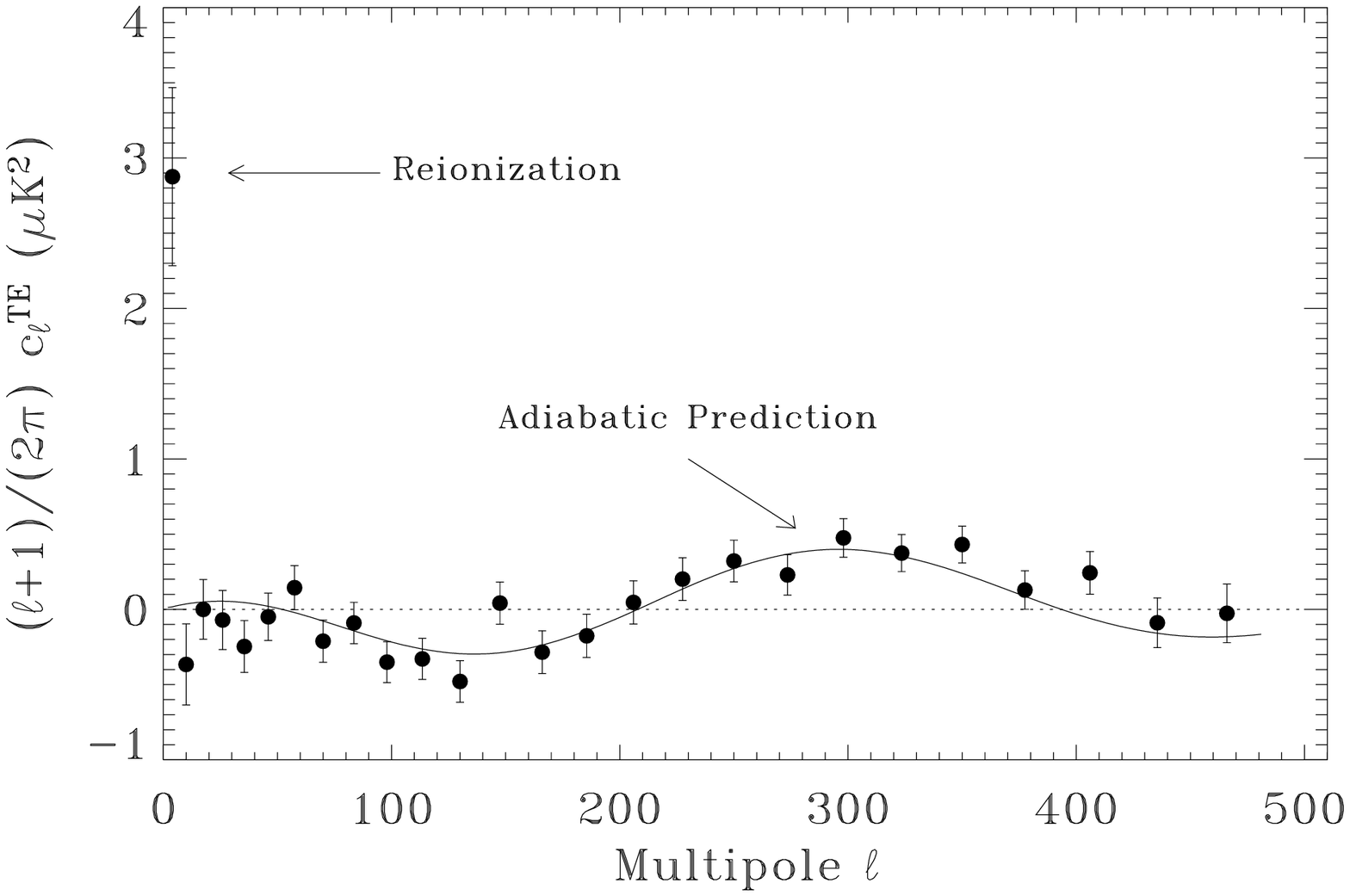}
\end{picture}
\end{picture}
\end{center}
\caption{Left: The TT power spectrum obtained from an optimal combination of 28 cross-power spectra. The solid line is the best fit model, and the gray area shows the cosmic variance error. Right: the temperature-polarization cross correlation power spectrum. The solid line is the prediction for the polarization signal from the TT data. The excess power at $\ell <10$ is due to early re-ionization (see \cite{Kogutetal03}).}
\end{figure}

\section{The Method}
The analysis method must match the level of accuracy and precision of the data.
 If the CMB sky is Gaussian, then the spherical harmonics coefficients $a_{\ell m}$ are normally distributed, but the angular power spectrum, $C_{\ell}$, is not. For large $\ell$ the Central Limit Theorem will make the Gaussian likelihood an increasingly good approximation, but the error-bars in the WMAP data are so small, that  previous likelihood approximations extensively used in the literature  are not suitable. We introduce a different form of the likelihood function that is accurate to better than 0.1\%. The likelihood function and its covariance matrix has been calibrated on 100,000 Monte Carlo realization of the sky as seen by WMAP \cite{Verdeetal03}.  
This calibration is also important because we can now compute the reduced $\chi^2$ for the ``best fit model'' and use it to compute the probability that the temperature pattern observed in the sky is a realization of the best fit model.

We perform a likelihood analysis for the cosmological parameters: at each point in parameter space (i.e., for a given set of cosmological parameters) a new model $C_{\ell}$ is computed  using  the publicly available code CMBfast 4.1 \cite{cmbfast}. 
Even with a supercomputer, at 2 seconds per evaluation, with a naive grid-based approach it would take a prohibitive amount of CPU time to perform the analysis for 6 or more parameters. 
We resort to Markov Chain Monte Carlo (MCMC): this approach has become the standard tool for CMB analyses \cite{CM, Cetal, LB, Kosowskyetal02}. In fact, for a flat reionized universe  we can evaluate the likelihood $\sim 120,000$ times in $<2$ days: this is adequate for finding the best fit model and reconstruction of the 1- and 2- $\sigma$ confidence levels for the cosmological parameters.
While more details are in \cite{Verdeetal03} and references therein, here we just remind the reader that a MCMC is similar to a  random walk in parameter space: at each step a new point is sampled in parameter space but  the density of steps (points in parameter space) is proportional to the {\it a posteriori} likelihood.
After an initial ``burn-in'' period, the chain has no dependence on the starting location.  
The important point to keep in mind about MCMC is that  a MCMC cannot tell us about regions of parameter space  where it has not been: it is very important that the chain achieves good mixing (i.e., moves rapidly in parameter space) and fully explore the likelihood surface.  Using an MCMC that has not fully explored the likelihood surface for determining cosmological parameters will yield {\it wrong} results, thus it is important to have a convergence criterion and mixing diagnostic. For any analysis of the WMAP data we strongly encourage  the use of a convergence criterion, for example we  use the method proposed by \cite{GR92}.

We speed up convergence and improve mixing by using the physical parameters of \cite{BE} and \cite{Kosowskyetal02} (e.g., physical cold dark matter density $\Omega_c h^2$ instead of $\Omega_c$ and the characteristic angular scale of the acoustic peaks $\Theta_A$ instead of the Hubble constant).

\subsection{WMAPonly results}
We first present the cosmological parameter constraints from WMAP data only. Then we consider external data sets (smaller scale CMB experiments and large scale structure power spectrum measurements) and compare these data to the predictions of the best fit model for WMAP data. We find that this extrapolation of WMAP predictions to low redshift and smaller scales works remarkably well. We thus proceed to perform a joint likelihood analysis of WMAP data with these external data sets.
Apart from the fact that the CMB seems to be Gaussian \cite{Komatsuetal03}, consistent with the  generic prediction of inflationary models, and that the Universe seems to have been reionized much earlier that previously though \cite{Kogutetal03}, the most striking result, probably, is that the ``vanilla'', flat LCDM model still fits the data amazingly well: 6 parameters  fit 1348 data-points (these are the power spectrum multipoles, one could be bolder and say that 6 parameters  fit $\sim 10^6$ data, these being the pixels!).
The parameters that best fit WMAP data are summarized in table 1.

Remarkably, this same model fits not only WMAP data but also other small scale CMB experiments and a host of other cosmological observations:
the Hubble constant constraint ($H_0=72\pm 5$) is in agreement with the Hubble key project \cite{Freedmanetal01} determination of $H_0=72\pm 3 \pm 7$ at the 68\% confidence level(C. L.); the age  determination of $13.4 \pm 0.3 Gyr$ is in good agreement with other age-determination  based on stellar ages; the baryon abundance $\Omega_b h^2=0.024 \pm 0.001$ implies a primordial deuterium abundance in good agreement with D/H determination from quasar absorption lines; the determination of amplitude of fluctuations parameterized by $\sigma_8$  (the r.m.s. fluctuations smoothed on a scale of $8 Mpc/h$) is in good agreement with clusters and weak lensing estimates.

\section{External data sets}
We also consider smaller scales CMB experiments  ACBAR \cite{Kuoetal02} and CBI \cite{Masonetal02, Pearsonetal02} on scales where they do not overlap with WMAP measurements and they are not affected by possible  SZ signal.
We  combine CMB data with measurement of the low redshift universe. Galaxy redshift surveys allow us to measure the galaxy power spectrum at $z \sim 0$ and observations of the Lyman $\alpha$ (Ly${\alpha}$) forest allows us to probe the dark matter power spectrum at $z \sim 3$.
We use the Anglo-Australian Telescope Two Degree Field Galaxy redshift Survey (2dFGRS;\cite{Collessetal01}),
and the linear matter power spectrum as recovered by \cite{Croftetal02} and \cite{Gnedinhamilton02} from Ly${\alpha}$ forest observations.
When analyzing the 2dFGRS survey we find that we need to model the non-linear evolution of the power spectrum and non-linear redshift space distortions, even when limiting the analysis to a regime traditionally considered to be still  well described by linear theory. We tested our modeling extensively on mock catalogs of the 2dFGRS. The goal of our modeling is to relate the shape and the amplitude of the observed power spectrum to that of the underling linear dark matter power spectrum as constrained by CMB data. Thus we use the information about the redshift space-distortion parameters and galaxy bias  as in \cite{Peacoketal01} and \cite{Verdeetal02} (for more details see \cite{Verdeetal03}).

For the matter power spectrum derived from    Ly${\alpha}$ forest observations we follow the approach on \cite{Gnedinhamilton02} and increase the error-bars where their analysis is not in agreement with that of \cite{Croftetal02}). We stress here that the addition of  Ly${\alpha}$ forest data appear to confirm trends seen in the other data sets and tighten cosmological constraints however our results do not rely on Ly${\alpha}$ data. More observational and theoretical work is still needed  to confirm the emerging consensus that the  Ly${\alpha}$ forest data traces the large scale structure (see L. Hui and U. Seljack talks)

\begin{figure}[htbp]
\label{fig:cls}
  \begin{center}
    \plotone{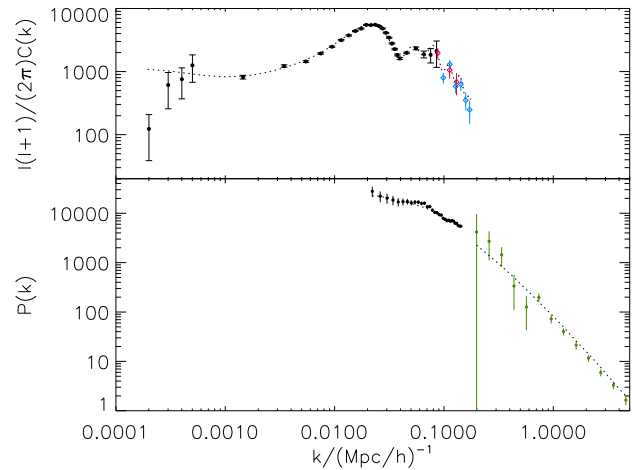}
    \caption{The data set, and the LCDM model best fit to WMAP data.   In the top panel WMAP data (black) CBI (red) ACBAR (blue) where the multipole $\ell$ has been converted to wavenumber $k$. In the bottom panel 2dFGRS (black) and Ly$\alpha$ (green). The extrapolation of this model to low redshift and to smaller scales fits the data remarkably well. For 2dFGRS we have used redshift space distortions parameters: $\beta=0.45$, $\sigma_v=380$ km/s, and bias parameter $b=1.13$.}
\end{center}
\end{figure}

The key cosmological parameters that are relevant for single field inflation  are: the constraint on the total matter-energy density of the universe relative to the critical density $\Omega_{tot}$, the spectral slope of the primordial power spectrum $n_s$ the running of the spectral index $dn/d\ln k$ and the tensor to scalar ratio $r$. These are reported in table 1 and in figure \label{fig:inflation}.

\begin{table*}[h]
\label{tab:WMAPLCDM}
\begin{center}
\begin{tabular}{|c c c c c|}
\hline
LCDM model & & & & \\
Parameter & WMAP & +HST &+SN&  \\
\hline
$\Omega_{tot}$ & $1.038^{+0.037}_{-0.04}$& $1.001\pm 0.018$ & $1.008\pm 0.017$ & \\  
\hline
Parameter ($\Omega_{tot}\equiv 1$) & WMAP & +CBI+ACBAR  &+2dFGRS & +Ly$\alpha$ \\
\hline
$n_s$ &$0.99\pm 0.04$& $0.97\pm 0.03$ & $0.97\pm 0.03$& $0.96\pm 0.02$\\
\hline
\hline
Running index model& ($r\equiv0$ & $\Omega_{tot}\equiv 1$)& &\\
Parameter & WMAP & +CBI+ACBAR &+2dFGRS&+Ly$\alpha$\\
$n_s$ $(k_0=0.05 Mpc^{-1})$&$0.93\pm 0.07$ & $0.91\pm 0.06$ & $0.93^{0.04}_{0.06}$ & $0.93\pm 0.03$\\
$d n_s/d\ln k$&$-0.047\pm 0.04$&$-0.055\pm 0.038$& $-0.031^{0.023}_{0.025}$&$-0.031^{0.016}_{0.017}$ \\
\hline
\hline
Running index Tensors&  & ($\Omega_{tot}\equiv 1$)& & \\
Parameter & WMAP &  &+2dFGRS& +2dFGRS+Ly$\alpha$\\
$n_s$ $(k_0=0.002 Mpc^{-1})$ &$ 1.20^{+0.12}_{-0.11}$&   &$1.18^{+0.12}_{-0.11}$ & $1.13\pm 0.08$     \\
$d n_s/d\ln k$ &$-0.077^{+0.050}_{-0.052}$ &   &$-0.075^{+0.044}_{-0.045}$ &$-0.055^{+0.028}_{-0.029}$ \\
$r$ $(k_0=0.002 Mpc^{-1})$(95\% constraint) & $<1.28 $ & $   $ & $<1.14$  & $<0.9$ \\
\hline
\hline
\end{tabular}
\end{center}
\end{table*}

\begin{figure*}
\begin{center}
\setlength{\unitlength}{1mm}
\begin{picture}(90,98)
\includegraphics{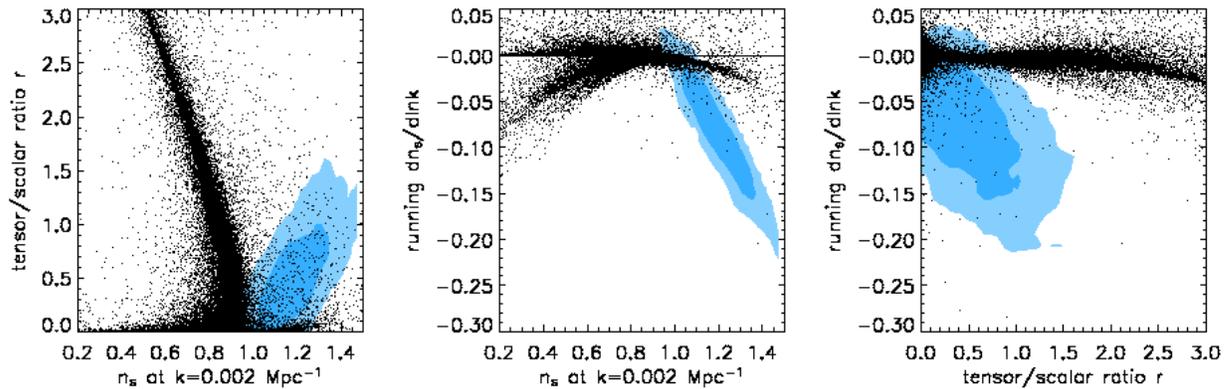}
\end{picture}
\caption{The parameter space spanned by viable slow-roll inflation models, with the WMAP+CBI+ACBAR+2dFGRS+Ly$\alpha$ 68\% confidence region in dark blue and 95\% confidence region in light blue (see \cite{Peirisetal03} for more deatils).}
\end{center}
\end{figure*}

\section{Results and conclusions}

I will present here 2 sets of conclusions: 
\begin{itemize}
\item{}1) for the pessimists I have shown how to pollute the clean CMB data with the messy {\it gastrophysics} of the low redshift probes of  large-scale structure. 
\item{}2) for the optimists, I have shown that all these different and independent data sets are in remarkable  agreement. Within the context of the inflation + LCDM paradigm cosmological parameters constraints obtained from these different datasets are consistent. The results obtained at $z\sim 1088$ can be extrapolated forward to $z\sim 0$ and describe a host of astronomical observations.

This gives us the confidence to extrapolate our results backwards, to well before $z\sim 1088$ and place constraints on Inflation, as presented in \cite{Peirisetal03} (see also   D. Spergel talk and  fig. 3).
\end{itemize}

For both optimists and pessimists I hope I have illustrated how it is possible to enhance the scientific value of CMB data from $z\sim 1088$ by combining it with measurements of the low redshift  universe.
When including these external data sets one should keep in mind that the underlying physics for these data sets is much more complicated and less well understood than for CMB data and especially  WMAP data. 
For both pessimists and optimists I hope I have highlighted the importance of 
improving the rigor of large-scale structure data analysis and of the  
interpretation of these observations,  and how crucial it is to quantify an d account for the level of systematic contamination.

\begin{small}
The data set, the WMAP temperature and temperature-polarization angular power spectra are available at http://lambda.gsfc.nasa.gov. If using these data set please refer to Hinshaw et al. (2003) for the TT power spectrum and Kogut et al. (2003) for the TE power spectrum. We also make available a subroutine that computes the likelihood for WMAP data given a set of $C_{\ell}$, please refer to Verde et al. (2003) if using the routine.
\end{small}



\section{Acknowledgments}

I thank my collaborators in this work: C. Barnes, C. Bennett (PI), M. Halpern, R. Hill, G. Hinshaw, N. Jarosik, A. Kogut, E. Komatsu, M. Limon, S. Meyer, N. Odegard, L. Page, H. Peiris, D. Spergel, G. Tucker, L. Verde, J. Weiland, E. Wollack, E. Wright. (the WMAP science team). The WMAP mission is made possible by the support of the Office of Space Science at NASA Headquarters.
LV is supported by NASA through Chandra Fellowship PF2-30022 issued by the Chandra X-ray Observatory center, which is operated by the Smithsonian Astrophysical Observatory for and on behalf of NASA under contract NAS8-39073. The WMAP mission is made possible by the support of the Office of Space Science at NASA Headquarters.
I thank the  2dFGRS consortium for help and discussion.

\end{document}